\documentclass{article}

\usepackage{amsmath,amssymb}
\usepackage{algorithm}
\usepackage{algpseudocode}
\usepackage{subfigure}
\usepackage{graphicx}
\usepackage{arxiv}

\usepackage[utf8]{inputenc} 
\usepackage[T1]{fontenc}    
\usepackage{hyperref}       
\usepackage{url}            
\usepackage{booktabs}       
\usepackage{amsfonts}       
\usepackage{nicefrac}       
\usepackage{microtype}      
\usepackage{lipsum}		

\title{A data-driven choice of misfit function for FWI using reinforcement learning}

\author{
  Bingbing Sun  \\
  Physical Sciences and engineering\\
  King Abdullah University of Science and Technology\\
  Thuwal, 23955, Saudi Arabia \\
  \texttt{bingbing.sun@kaust.edu.sa} \\
   \And
  Tariq Alkhalifah \\
  Physical Sciences and engineering\\
  King Abdullah University of Science and Technology\\
  Thuwal, 23955, Saudi Arabia \\
  \texttt{tariq.alkhalifah@kaust.edu.sa} \\
}

\begin{document}
\maketitle

\begin{abstract}
In the workflow of Full-Waveform Inversion (FWI), we often tune the parameters of the inversion to help us avoid cycle skipping and obtain high resolution models. For example, typically start by using objective functions that avoid cycle skipping, like tomographic and image based or using only low frequency, and then later, we utilize the least squares misfit to admit high resolution information. We also may perform an isotropic (acoustic) inversion to first update the velocity model and then switch to multi-parameter anisotropic (elastic) inversions to fully recover the complex physics. Such hierarchical approaches are common in FWI, and they often depend on our manual intervention based on many factors, and of course, results depend on experience. However, with the large data size often involved in the inversion and the complexity of the process, making optimal choices is difficult even for an experienced practitioner. Thus, as an example, and within the framework of reinforcement learning, we utilize a deep-Q network (DQN) to learn an optimal policy to determine the proper timing to switch between different misfit functions. Specifically, we train the state-action value function (Q) to predict when to use the conventional L2-norm misfit function or the more advanced optimal-transport matching-filter (OTMF) misfit to mitigate the cycle-skipping and obtain high resolution, as well as improve convergence. We use a simple while demonstrative shifted-signal inversion examples to demonstrate the basic principles of the proposed method. 
\end{abstract}


\section{Introduction}
Considering the high nonlinearity of Full Waveform Inversion (FWI), a hierarchical approach is common practice. Developing such a strategy is a daunting task considering the size of the data, the realities of how the data were acquired (limited band and aperture), and the physical assumptions we impose on the model. With respect to the choice of the proper objective function, recent advances admitted reasonably cycle-skipping free misfit functions such as the matching filter misfit, optimal transport (OT) \cite{opt2} function or a combination of them, i.e., the optimal transport of the matching filter misfit (OTMF) \cite{MF_OTMF} . Unlike the L2-norm misfit, which is a local comparison, these advanced misfit functions seek global comparisons between the predicted and measured data, and thus, can avoid the cycle-skipping. However, we often still need to switch to the L2 norm for higher resolution models when the data are less cycle-skipped and safe for local comparisons. In reality, we need to carefully QC of the data matching to determine the optimal time to switch. Besides, the probability of cycle-skipping varies for different offsets. It would be ideal to use different misfit functions for different traces to accommodate their specific cycle-skipping probabilities. 
In principle, we can formulate this problem as that in each iteration, given the predicted and measured data, we try to make a decision (action) to choose between the L2 norm misfit and a cycle-skipping free misfit such as the OTMF and the consequence of such action seeks a better fitting of the data in a long time horizon ( running over many iterations). Mathematically, it is considered as a Marko decision process and it is well studies in the field of statistics and machine learning. In this paper, based on the concept of reinforcement learning, we train a Deep Q network (DQN) \cite{DQN} to achieve fast convergence by learning the optimal choice of objective functions over FWI iterations.  

Reinforcement learning (RL) is one of three basic machine learning paradigms, alongside supervised learning and unsupervised learning. It is a potential algorithm towards true artificial intelligence. It differs from supervised learning that it does not require labels given by input/output pairs. Instead, the focus of RL is in finding a balance between exploration (of uncharted territory) and exploitation (of current knowledge). Recently, RL based algorithms demonstrated its potential in solving complex problems which is extremely difficult for conventional machine learning algorithms, e.g., AlphaGo beats the top human Go players while the Alphastar achieves the Grandmaster level at playing the Starcraft games \cite{Silver2017,Vinyals2019}. In this paper, we share our first attempt to use RL to automatically selecting a misfit function in full-waveform inversion. We start with a brief review of the OMTF misfit function used here and then develop the method for misfit function selection using DQN. At last, we demonstrate our method using a time-shifted signal example.

\section{A robust misfit function by optimal transport of the matching filter (OTMF)}
\label{sec:headings}
The conventional L2-norm misfit function seeks a local point-wise comparison between the predicted data $p(t)$ and the measured data $d(t)$:
	\begin{equation}
		J_{L2} = \frac{1}{2} ||p(t)-d(t)||_2^2.
	\end{equation}
	\cite{MF_OTMF} introduce the optimal transport of the matching filter (OTMF) misfit for FWI. In OTMF, a matching filter would be computed first by deconvolving of the predicted data with the measured data:
	\begin{equation}
		d(t)*w(t)=p(t),
	\end{equation}
	where $*$ denotes the convolution operation. After a proper precondition of the resulting matching filter to fulfill the requirements for a distribution,  we minimize the Wasserstein $W_2$ distance between the resulting matching filter and a target distribution given by, e.g., a Dirac delta function:
	\begin{equation}
	J_{OTMF} = W_2^2(\tilde{w}(t),\delta(t)),
	\end{equation}
	where $\tilde{w}(t)={w^2/}{||w||_2^2}$, $W_2$ denotes the Wasserstein distance \cite{opt2}. The resulting OTMF misfit in Equation 3 can overcome the cycle-skipping effectively as demonstrated by \cite{SUN_OTMF_EAGE}.
	
\section{Automatic misfit function selection using Deep Q network (DQN) }
	We first provide a mathematical background summary for the Markov decision process (MDP), the deep Q network (DQN) and the RL techniques. We adopt the standard MDP formalism. An MDP is defined by a tuple $<S,A,R,P,\gamma>$, which consists of a set of states $S$, a set of actions $A$, a reward function $R(s,a)$, a transition function $P(s'|s,a)$ and a discount factor $\gamma$. For each state $s \in S$, the agent takes an action $a \in A$. Upon taking this action, the agent receives a reward $R(s,a)$ and reaches a new state $s'$ as a result of the action, determined from the probability distribution $P(s'|s,a)$. In RL, we try to learn a policy $\pi$, specified for each state which action the agent will take. The goal of the agent is to find such policy $\pi$ mapping states to actions that maximizes the expected discounted total reward over the agent's lifetime. Such long time expected reward is formulated as the action-value function (Q) : 
\begin{equation}
Q^{\pi}(s,a) = \mathbb{E}^{\pi}\left[\gamma^t R(s_t,a_t)\right], 
\end{equation}
where $\mathbb{E}^{\pi}$ is the expectation over the distribution of the admissible trajectories $(s_0,a_0,s_1,a_1,...)$ obtained by executing the policy $\pi$ starting from $s_0=s$ and $a_0=a$. 
There are many algorithms developed in RL to learn the policy $\pi$. DQN is a popular method for dealing with a discrete action space. It only learns the Q function and the optimal policy $\pi^*$ can be derived from the learned Q function directly:
\begin{equation}
	\pi^*(a|s) = \max_{a'} Q(s,a').
\end{equation}
Equation 5 is intuitive to understand that the best action for each state should give the largest Q value for that state. In order to learn the Q function, we take a single move from current state to next one and see what reward R we can get. This admits a one-step look ahead:
\begin{equation}
Q(s_{t},a_t) = r_{t+1}+\gamma \max_{a'} Q(s_{t+1},a').
\end{equation}
In order to stabilize the learning process, we keep track of another target Q function: $Q'$. Thus, the loss function of DQN in training is the time difference (TD) error between the Q function and its target :
\begin{equation}
\text{Loss}= \frac{1}{2}\left[ r_{t+1}+\gamma \max_{a'} Q'(s_{t+1},a')-Q(s_{t},a_t)\right]^2
\end{equation}
For training efficiency, we save the transition $[ s_t, a_t, r_t, s_{t+1}]$ in a replay buffer and reuse these datasets for training (It is referred as experience replay in RL).

Exploration plays an important role in RL. Exploration provides the agent with the ability to expand his knowledge when interacting with the environment. The $\epsilon$-greedy exploration strategy randomly choses the action given a probability $\epsilon$:
\begin{equation}
	a_t = \begin{cases}
		 a^*_t                      & \text{with probability} \quad1-\epsilon, \\
		 \text{random action} & \text{with probability} \quad \epsilon, 
		 \end{cases}
\end{equation}
where $a^*_t$ is related to the optimal policy from equation 5. We start with a large $\epsilon$ and gradually reduce it during the training. 
Another important aspect related to RL is the reward, the decision of the form for the reward is problem-specific, and it may affect the training significantly. Algorithm 1 shows a typical DQN flow with experience replay and $\epsilon$-greedy exploration policy.
\begin{algorithm}
\caption{Deep Q learning}
\begin{algorithmic}[1]
\State Initialize replay memory D to capacity N
\State Initial action-value function Q with random weights $\theta$
\State Initialize target action-value function $\hat{Q}$ with weights $\hat{\theta}=\theta$ 
\For{${episode=1,M}$}
\For{${t=1,T}$}
\State With probability $\epsilon$, select a random action $a_t$
\State otherwise select $a_t = \max_{a'}Q(s_t,a;\theta)$
\State execute action $a_t$ and observe reward $r_{t+1}$ and the next state $s_t$
\State store transition $(s_t,a_t,r_t,s_{t+1})$ in D
\State Sample random minibatch of transitions $(s_t,a_t,r_t,s_{t+1})$ from D
\State Set $y_j = r_{j+1} + \gamma \max_{a'} \hat{Q}(s_{j+1},a;\hat{\theta})$
\State Perform a gradient descent step on $\frac{1}{2}(y_j-Q(s_j,a_j;\theta))^2$ for updating $Q$ paramter $\theta$
\State Every C steps reset $\hat{Q}=Q$
\EndFor
\EndFor
\end{algorithmic}
\end{algorithm}

It is straightforward to adapt DQN to our misfit function selection problem, i.e., select between the L2 norm and the OTMF misfit. Considering a one dimensional FWI problem, the state in RL would be the predicted and measured data:
\begin{equation}
	s_t = (p_t,d_t),
\end{equation} 
where $p_t$ and $d_t$ is a single trace of the data in the time domain at iteration step $t$. The Q function will have such a state as input and it will output two values determining whether we use the L2 norm misfit function or the OTMF misfit function. We will also incorporate the $\epsilon$-greedy exploration policy, i.e., we will random choose between the L2 norm and the OTMF misfit with probability $\epsilon$.  For the reward, we can define it as the negative of the normalized L2 norm of the model difference, or the negative of the normalized L2 norm of the data residuals as another option.  
\begin{equation}
r_t = -\frac{||m_{\text{true}}-m_{t}||_2^2}{m_{\text{true}}^2} \quad \text{or} \quad r_t = -\frac{||p_{\text{true}}-p_{t}||_2^2}{p_{\text{true}}^2}
\end{equation}
We should keep in mind that unlike in FWI, here though we use a L2 norm of the data difference to formulate the reward in the RL training, it will not be an issue. Because the Q function we try to fit in Equation 4 seeks a long time expected reward (over many iterations). This means that the best policy learnt will always give fast convergence with less accumulated L2 norm of the data residuals throughout the inversion process. 

\section{Results}

In this example, we try to optimize a single parameter, i.e., the time shift between signals. An assumed forward modeling produces a shifted Ricker wavelet, using the formula
\begin{equation}
	F(t;\tau,f) = \left[1-2\pi^2f^2(t-\tau)^2\right]e^{-\pi^2f^2(t-\tau)^2}
\end{equation}
where $\tau$ is the time shift and $f$ is the dominant frequency. The modeling equation given by equation 11 is a simplified version of a PDE based simulation. The reward we use for the training is the normalized L2-norm data residuals (the second formula in Equation 10). 
	In this example, the data are discretized using $nt=200$ samples with a time sampling $dt=0.01$ s. We use direct connected network (DCN) for the Q function. We use one hidden layer for the DCN of size $nt$. The Q network will output two scalar values representation the Q for the L2 norm and the OTMF. We set the initial value of $\epsilon$ to be 0.90 and drop it exponentially to 0.05 at the end. Using a 3 Hz peak frequency wavelet, we randomly generate the initial and true time-shifts between 0.4 s and 1.2 s. In each episode (one full run of the inversion), we iterate for 12 iterations. We run ten thousand episodes for training, and we update the Q network based on equation 7 at every iteration. The batch size is set to be 128, i.e., we randomly fetch 128 tuples of $(s_t,a_t,r_t,s_{t+1})$ for updating the Q function.	
	Figure 1a shows the Loss of equation 7 over episodes (the curves in Figure 1 has been smoothed with a moving average over 100 episodes). Its convergence demonstrates the success of the RL training. Figure 1b is the accumulated reward over episodes, and its increasing value further indicates the learnt policy improved and can achieve fast convergence with higher reward throughout the training.  
	In order to further understand the trained Q function, we plot the Q value for different time shifts (we set the measured data with time-shift 0.5 s and scan the Q function over the predicted data with time-shift varying from 0.5 to 1.1 s). We plot the Q function over the relative time-shift between the predicted and measured data in Figure 2a. Figure 2b denotes the action that will be taken based on the learnt Q function (0 for L2 norm, 1 for OTMF). We can see that if the relative time shift is smaller than approximate 0.15 s, the Q value for the L2 norm is larger than the OTMF, suggesting apply the L2-norm misfit function. Otherwise, the learnt Q function would suggest to use OTMF to avoid the cycle-skipping. Note the switch point at 0.15 s is consistent with the half cycle of the 3 Hz peak frequency we used in training. However, this number is fully determined from the data itself in the framework of reinforcement learning. 
	\section{Conclusions}
		In the framework of Reinforcement Learning, we trained a Deep Q network (DQN) to select a misfit function for FWI. We use the time-shift inversion example to demonstrate the basic principle of our method. The resulting trained network managed to use the data to determine the appropriate objective function to achieve convergence.
\begin{figure*}
\centering
\subfigure[]{\includegraphics[width=0.40\textwidth]{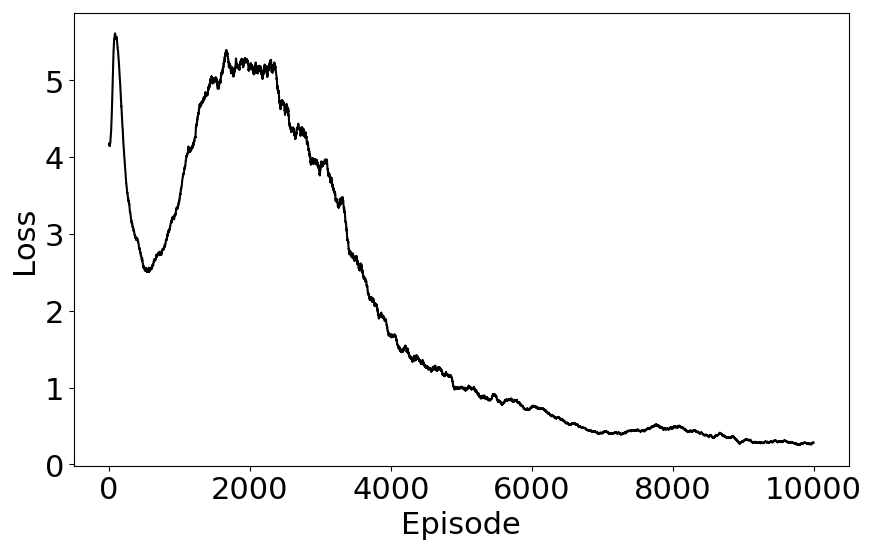}}
\subfigure[]{\includegraphics[width=0.40\textwidth]{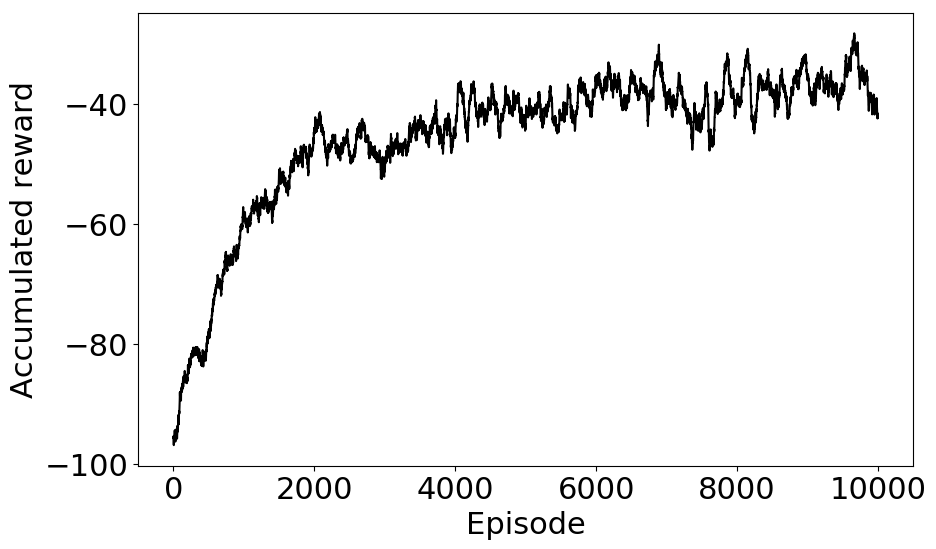}}
\caption{
a) The loss value of equation 7 over episodes; b) The accumulated reward over episodes. }
\end{figure*}

\begin{figure*}
\centering
\subfigure[]{\includegraphics[width=0.40\textwidth]{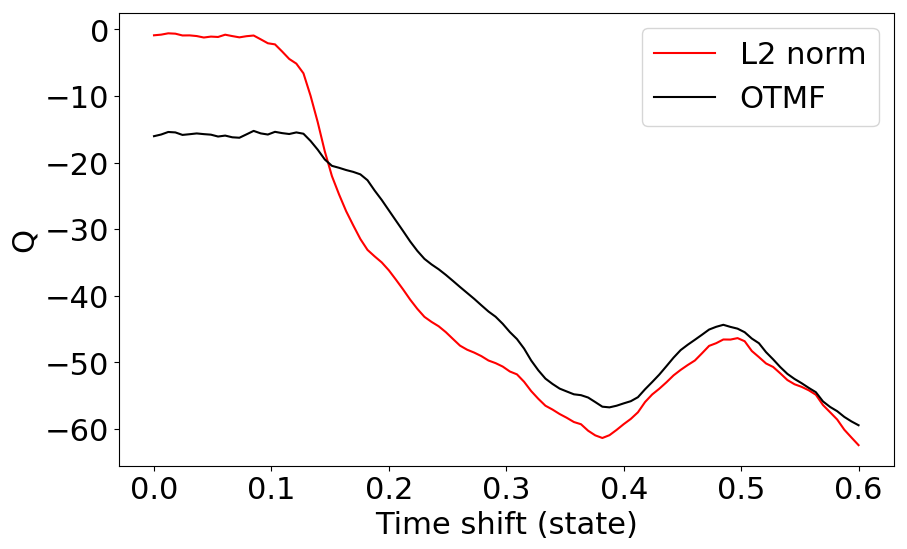}}
\subfigure[]{\includegraphics[width=0.40\textwidth]{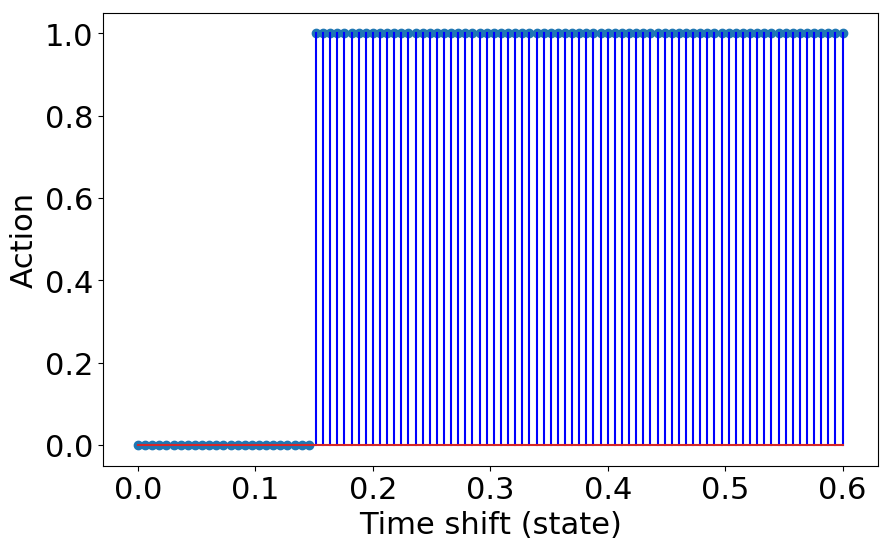}}
\caption{
a) The Q function for different states (time shift); b) The actions the policy takes for different states (0 for L2 norm, 1 for OTMF). }
\end{figure*}
\bibliographystyle{unsrt}  
\bibliography{mypaper}
\end{document}